# Applicability of Gaussian Noise Approximation for Optically Pre-Amplified DPSK Receivers with Balanced Detection

**Xiupu Zhang, Zhenqian Qu**

*Department of Electrical and Computer Engineering,*

*Concordia University, Montreal, Quebec, CANADA*

*Tel: 514 848 2424 ext.4107, Fax: 514 848 2802,*

*E-mail: xzhang@ece.concordia.ca*

**Guodong Zhang**,

*AT&T, 200 Laurel Avenue, Middletown, NJ 07748, USA*

**Guangxue Yang**

*School of Electrical and Electronic Engineering,*

*Harbin University of Science and Technology, CHINA*

**Abstract:** This letter presents a comparison of exact probability density function with the Gaussian noise approximation in optically pre-amplified DPSK receivers with optical Mach-Zehnder interferometer demodulation (MZI) and balanced detection, including the impact of phase noise. It is found that the Gaussian noise approximation significantly over-estimates ASE-ASE beat noise in DPSK receivers with balanced detection particularly when phase noise is negligible, compared to IM/DD receivers, ASE- amplified spontaneous emission. However, the Gaussian noise approximation is still applicable for DPSK receivers with balanced detection and the measured 3-dB advantage is predicted by the Gaussian noise distribution.

***Indexing Terms***: Optical fiber communication, optical receiver, optical modulation, differential phase shift keying, wavelength division multiplexing.

**I. Introduction**

Differential phase shift keying (DPSK) with optical Mach-Zehnder interferometer (MZI) demodulation and balanced detection has gained much attention since several advantages over intensity modulation and direct detection (IM/DD) have been discovered [1-5]. Particularly, the requirement of optical signal to noise ratio (OSNR) is reduced by 3 dB, or the transmission distance is doubled [1-5] for the same system performance by using DPSK/MZI receivers with **b**alanced **d**etection (DPSK/MZI-BD), compared to DPSK/MZI receivers with **s**ingle-port **d**etection (DPSK/MZI-SD) or IM/DD (*DPSK/MZI-SD is equivalent to IM/DD if both optically pre-amplified*). There has been much discussion on how to explain the 3-dB advantage origin theoretically [6-9]. One origin is attributed to the fact that the Gaussian noise approximation is not adequate for DPSK/MZI-BD [6-8] and the other is given by the fact that the calculation method of bit error ratio (BER) is different from the conventional method [9]. In optically pre-amplified DPSK/MZI receivers, there mainly exist two noise contributions: amplified spontaneous emission (ASE) noise, which is added into the signal linearly (linearly additive ASE noise); and phase noise, which mainly consists of two parts: one part induced by ASE orthogonal component (linear phase noise), and the other part induced by nonlinear Kerr interaction between ASE noise and signal (nonlinear phase noise) [10-12]. If only considering the linearly additive ASE noise, i.e. signal-ASE beat noise and ASE-ASE beat noise, the Gaussian noise approximation cannot predict the 3-dB advantage [6-9] by using the definition of $BER = \frac{1}{2}\left[\int_{I_{th}}^{\infty} f_0(x)dx + \int_{-\infty}^{I_{th}} f_1(x)dx\right]$, where $f_1(x)$ and $f_0(x)$ are the Gaussian probability density functions (pdf's) of bits "1" and "0", and $I_{th}$ is the optimal decision threshold. In [6-8], the 3-dB advantage was predicted by using the exact pdf's based on the above BER. However, we will show that the 3-dB advantage predicted in [6-8] is not the measured 3-dB in [2-5]. It was shown that the noise statistics of differential phase noise (or phase noise difference) is well approximated by the Gaussian distribution [11-12]. If the effects of linearly additive ASE noise and differential phase noise both are taken into account, it was reported that inner tails of the pdf's of bits "1" and "0" are moved up due to the phase noise, and the 3-dB advantage finally is vanished if the phase noise is increased to some extent [13].

It was well established that the linearly additive ASE noise statistic (i.e. signal-ASE beat noise and ASE-ASE beat noise) follows the Chi-square distribution in IM/DD [14-15]. However, because the signal-ASE beat noise is usually dominating and has the Gaussian distribution, the Gaussian noise approximation has been widely used and provides a fairly good estimation of BER [14-16]. The only difference of between DPSK/MZI and IM/DD receivers is that an optical MZI is inserted before optical photodiodes, and the optical MZI is a special optical filter. Hence, the signal detection and ASE processing in DPSK/MZI receivers are almost the same as in IM/DD receivers. We could expect that there is no big difference in noise statistics between DPSK/MZI and IM/DD receivers except that DPSK/MZI receivers are not immune to phase noise. In this letter, the exact pdf's of noise statistics in DPSK/MZI-BD are investigated including the impact of phase noise. The applicability of the Gaussian noise approximation for DPSK/MZI-BD is discussed.

## II. Theory

DPSK/MZI-BD consists of the following components in series; an optical pre-amplifier, an optical filter, an ideal optical MZI demodulator, balanced photodiodes and an electrical filter. For the ideal DPSK i.e. with no phase error, when bit "1" is received the signal completely presents at the constructive port, and only ASE noise will appear at the destructive port; and vice versa for bit "0"". For DPSK-BD with phase error, the currents for bits "1" and "0" are given approximately by

$$I_1(t)/R \approx P_s \cos(\Delta\Phi) + E_{s+} n_+^*(t) + E_{s+}^* n_+(t) + |n_+|^2 - |n_-|^2 \quad (1a)$$

$$I_0(t)/R \approx -P_s \cos(\Delta\Phi) - E_{s-} n_-^*(t) - E_{s-}^* n_-(t) - |n_-|^2 + |n_+|^2 \quad (1b).$$

$E_{s+}$ ($n_+(t)$) and $E_{s-}$ ($n_-(t)$) denote output electric fields of signal (ASE noise) at the constructive and destructive ports; and $P_s$ is the average signal power. R denotes the responsivity of the photodiodes. $\Delta\Phi$ denotes differential phase noise (or phase noise difference). The second and third terms represent signal-ASE beat noise and the last two

X. Zhang, et al.,                          Applicability….

terms are ASE-ASE beat noise. The pdf for the first four terms in (1a) is given by [14-15]

$$f_+(x|\Delta\Phi) = \frac{M}{\overline{I}_+}\left(\frac{x}{\overline{I}_1}\right)^{\frac{M-1}{2}}\exp\left(-M\frac{x+\overline{I}_1}{\overline{I}_+}\right)I_{M-1}\left(\frac{2M\sqrt{x\overline{I}_1}}{\overline{I}_+}\right) \quad x\geq 0, \qquad (2),$$

where $I_{M-1}(\ )$ - modified Bessel function, $M = B_o/B_e$, $\overline{I}_+ = 2RN_{ASE}B_+$ -the average current induced by ASE noise at the constructive port, and $\overline{I}_1 = RP_s\cos\Delta\Phi$ - the signal decision current of bit "1", $B_o$ -optical noise bandwidth before the MZI, $B_+ = B_o/2 + \sin(\pi B_o T_b)/(2\pi T_b)$ -equivalent optical noise bandwidth at the constructive port, $B_e$ - electrical receiver noise bandwidth, $T_b$ -bit period, and $N_{ASE}$ - the power spectral density of ASE noise at one polarization state. The last term in (1a) i.e. $|n_-(t)|^2$, which is from the destructive port, has the pdf given by [14-15]

$$f_-(x) = \left(\frac{M}{\overline{I}_-}\right)^M \frac{x^{M-1}}{\Gamma(M)}\exp\left[-M\frac{x}{\overline{I}_-}\right] \quad x\geq 0, \qquad (3),$$

where $\overline{I}_- = 2RN_{ASE}B_-$ - the average current induced by ASE noise at the destructive port, $B_- = B_o/2 - \sin(\pi B_o T_b)/(2\pi T_b)$ - equivalent optical noise bandwidth at the destructive port, and $\Gamma(M)$ -Gamma function. Thus, the exact pdf for (1a) or bit "1" can be computed with (for $w = u - v$, the pdf of $w$ is given by $f_w(x) = \int f_u(x+y)f_v(y)dy$ [17] if $u$ and $v$ independent, $f_u(\ )$ and $f_v(\ )$ - pdf's of $u$ and $v$)

$$f_1(x|\Delta\Phi) = \int_0^\infty f_+(x+y|\Delta\Phi)f_-(y)dy \qquad (4),$$

where $f_+(x+y|\Delta\Phi)$ and $f_-(y)$ are given by (2) and (3). Considering phase noise, the total pdf for bit "1" is obtained by

$$f(x) = \int_{-\infty}^\infty f_{\Delta\Phi}(\Delta\Phi)f_1(x|\Delta\Phi)d\Delta\Phi = \int_{-\infty}^\infty f_{\Delta\Phi}(\Delta\Phi)d\Delta\Phi\int_0^\infty f_+(x+y|\Delta\Phi)f_-(y)dy \qquad (5),$$

where $f_1(x|\Delta\Phi)$ is given by (4), and $f_{\Delta\Phi}(\Delta\Phi)$ is the pdf of $\Delta\Phi$, which is well approximated by the Gaussian distribution [11-12].

**III. Comparison of Exact and Gaussian pdf's**

We first consider the pdf's for the case of *no phase noise*. The currents from (1) become

$$I_1(t)/R = P_s + E_{s+}n_+^*(t) + E_{s+}^*n_+(t) + |n_+|^2 - |n_-|^2 \quad (6a),$$

$$I_0(t)/R = -P_s - E_{s-}n_-^*(t) - E_{s-}^*n_-(t) - |n_-|^2 + |n_+|^2 \quad (6b).$$

Compared to IM/DD, one difference is that there is an additional term in (6), i.e. ASE-ASE beat noise from the destructive port for bit"1" and constructive port for bit "0". Fig.1(a) depicts the exact pdf's (solid) calculated by (4). The Gaussian approximation with the variance of $\sigma_1^2 = 2RN_{ASE}I_sB_e + R^2N_{ASE}^2(B_o - B_e)B_e$ is also displayed for comparison (dashed). It is shown that the exact pdf's are not symmetrical to the current mean; and the Gaussian approximation over-estimates the inner tails of the pdf's and thus BER accordingly. However, if the signal-ASE beat noise is only considered in (6) i.e. ignoring the last two terms in (6), the pdf's given by (4) become the Gaussian distribution [14-15], the same as bit "1" in IM/DD. Thus, the asymmetry of the exact pdf's is totally attributed to the ASE-ASE beat noise. Therefore, *the Gaussian noise approximation only over-estimates the ASE-ASE beat noise in DPSK/MZI-BD, similar to IM/DD*. The following parameters have been used in Fig.1(a); bit rate of 43 Gb/s, optical pre-amplifier with gain of 35 dB and noise figure of 5 dB, optical noise bandwidth of $B_o$=100 GHz; electrical bandwidth of $B_e$=33 GHz, R=1, and optical signal power of -30 dBm. The noise figure of 5 dB is used for enhancing ASE-ASE beat noise.

Now we consider the case of phase noise included. Fig.1(b) shows the exact pdf's (solid) calculated by (5) for the case of $\Delta\Phi$ having a standard deviation of 0.25 radians. The calculated pdf's are in good agreement in shape with the measured [13]. For comparison, $f_1(x|\Delta\Phi)$ in (5) is assumed the Gaussian distribution, and the total pdf's by (5) are also shown in Fig.1(b) (dashed). For this case, the over-estimation of inner tails of the pdf's is significantly reduced by the Gaussian approximation. In (1) if only considering the phase noise, we have

$$I_1(t) = RP_s \cos(\Delta\Phi) \qquad (7a),$$

$$I_0(t) = -RP_s \cos(\Delta\Phi) \qquad (7b),$$

compared to DPSK/MZI-SD,

$$I_1(t) = \frac{1}{2}RP_s[1+\cos(\Delta\Phi)] \Rightarrow 2I_1(t) - RP_s = RP_s \cos(\Delta\Phi) \qquad (8a),$$

$$I_0(t) = \frac{1}{2}RP_s[1-\cos(\Delta\Phi)] \Rightarrow 2I_0(t) - RP_s = -RP_s \cos(\Delta\Phi) \qquad (8b).$$

Comparison of (7) and (8) has shown that the balanced detection has the same performance as the single-port detection if the phase noise is only considered, which proves the observation in [13], of which the 3-dB advantage is vanished when the phase noise is increased to some extent.

**IV. Discussion of Gaussian noise approximation**

For the case of no phase noise, the Gaussian noise approximation is the worst case, the largest over-estimation of the inner pdf tails induced. Therefore, DPSK/MZI-BD for the case of *no phase noise* is only discussed here. In order to understand the impact of the Gaussian noise approximation, we calculate the cumulative probability (CP) by two methods.

Method #1: $CP = \frac{1}{2}\left[\int_{I_{th}}^{\infty} f_0(x)dx + \int_{-\infty}^{I_{th}} f_1(x)dx\right]$ is shown in Fig.2 with the exact (solid) and Gaussian (dashed) pdf's for DPSK/MZI-BD ($I_{th}=0$) and IM/DD (the decision currents $\bar{I}_1 = 2RP_s$ for bit "1" and $\bar{I}_0 = 0$ for bit "0" assumed). The CP is considered as BER in [6-8]. Fig.2 shows that the ~3-dB advantage or improvement is predicted based on the exact pdf's by comparing DPSK/MZI-BD with IM/DD, rather than ~1 dB by the Gaussian approximation. *Since the exact pdf's of bits "1" and "0" become Gaussian if the ASE-ASE beat noise is ignored (see (6)), the CP for both DPSK/MZI-BD and IM/DD become the same (details also in [9]) (CP almost overlapped with the dashed curve in Fig.2).* Therefore, the ~3-dB advantage in Fig.2 is attributed to the ASE-ASE beat noise. We have verified that the advantage is still kept ~3 dB for $B_o$ of down to $B_oT_b \approx 0.7$ (without considering signal distortion by filtering) and is decreased with the increase of $B_o$. This is in excellent agreement with in [6-8] (In [6-8], the ~3-dB advantage is quickly vanished with the increase of $B_o/B_e$. This suggests that the ~3-dB advantage is induced by ASE-ASE beat noise because the signal-ASE beat noise is independent of $B_o$). Because the ~3-dB advantage in Fig.2 is due to ASE-ASE beat noise, this ~3-dB advantage cannot be used for doubling the transmission distance or reducing the OSNR requirement of 3-dB. In other words, the performance of DPSK/MZI-BD and IM/DD ultimately becomes identical if CP by method #1 is BER. On the other hand, the measured advantage is typically in the range of 3-4 dB [1-5], and thus the measured is beyond the predicted. The measured 3-dB advantage is also interpreted by signal constellation [1]. In Fig.3, we plot the signal constellations for the single-port and balanced detections. The distance of bits "1" and "0" electric fields in DPSK/MZI-SD is assumed $x$, and then the distance is $\sqrt{2}x$ in DPSK/MZI-BD. Thus, the inherent 3-dB advantage in intensity is obtained by use of DPSK/MZI-BD. In other words, the 3-dB advantage or improvement is directly induced by the signal itself. Physically, the inherent 3-dB advantage can be explained as follows. In DPSK/MZI-SD or IM/DD (bit "0" always has zero decision current), if the decision current of bit "1" becomes zero due to some reasons, bits "1" and "0" are not distinguishable and thus errors occur; on the contrary, bits "1" and "0" are still distinguishable in DPSK/MZI-BD because bits "0" has non-zero decision current. Therefore, the ~3-dB advantage shown in Fig.2 is not measured 3-dB in [1-5]. It is worth to emphasize that the ~3-dB advantage in Fig.2 from the ASE-ASE beat noise in the two ports is only obtained theoretically by *ideally* balanced detection.

Method #2: $CP = \frac{1}{2}\{\text{Prob}[I_1(t) < I_0(t) | \text{bit} = 1] + \text{Prob}[I_0(t) > I_1(t) | \text{bit} = 0]\}$, which is considered as BER in [9], is discussed below. The relationship between the two methods is given by $\bar{P}|_{\#2}(\text{dBm}) = \bar{P}|_{\#1}(\text{dBm}) - 3(\text{dB})$ for a given CP, $\bar{P}|_{\#1}$ and $\bar{P}|_{\#2}$ - optical receiver sensitivity from methods #1 and #2. Thus, the advantage of ~6 (4) dB is obtained with the exact (Gaussian) pdf's (3 dB due to signal-ASE beat noise, and the left ~3 (1) dB due to ASE-ASE beat noise for our setting). Thus, DPSK/MZI-BD ultimately outperforms

DPSK/MZI-SD or IM/DD by exact 3 dB. Because this 3-dB advantage is not due to ASE-ASE beat noise and from the signal itself, the convergence of method #2 with the signal constellation is obtained. Therefore, the calculated CP by method #2 is BER for DPSK/MZI-BD, and the physical explanation of BER by method #2 will be given elsewhere.

**V. Conclusions**

We have compared the exact pdf's of noise statistics in optically pre-amplified DPSK/MZI-BD including phase noise with the Gaussian noise approximation. It is shown that the Gaussian noise approximation only induces a larger over-estimation of ASE-ASE beat noise in DPSK/MZI-BD than in IM/DD. The partial cancellation of ASE-ASE beat noise by *ideally* balanced detection can induce ~3-dB improvement of receiver sensitivity, predicted with the exact pdf's, rather than ~1 dB with the Gaussian approximation. However, we have found that this ~3-dB improvement is not experimentally measured 3-dB. The measured 3-dB advantage is predicted by the Gaussian approximation if the correct BER calculation is used (i.e. method #2). Consequently, the Gaussian noise approximation is still applicable for DPSK/MZI-BD as in IM/DD. Besides, we have shown that DPSK/MZI-BD has the same upper limit of phase noise as DPSK/MZI-SD. Thus, it is confirmed that the 3-dB advantage will be vanished if the phase noise is increased to the upper limit.

**Figure Captions:**

1. The pdf's of noise statistics for bits "1" and "0": (a) without phase noise; (b) phase noise included. The exact (solid), and Gaussian (dashed) pdf's.

2. Cumulative probability for DPSK/MZI-BD with the exact (solid) and Gaussian approximated (dashed) pdf's, and IM/DD with the exact (solid with dots) and Gaussian approximated (dashed with dots) pdf's. The parameters are the same as in Fig.1(a) except for noise figure of 3 dB.

3. Signal constellations for DPSK/MZI receivers with the single-port and balanced detections. The DPSK/MZI-SD is equivalent to IM/DD if both optically pre-amplified.

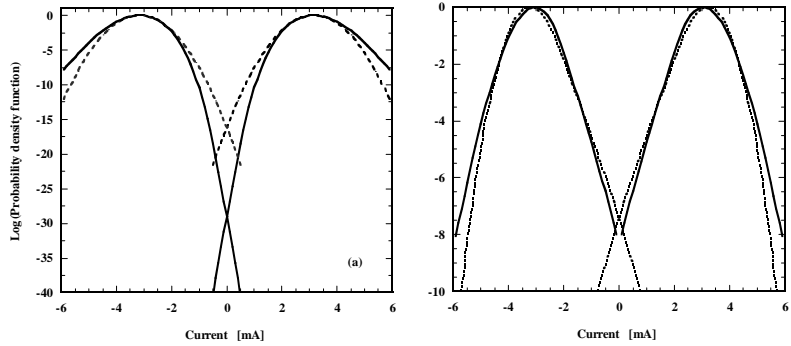

**Fig.1**

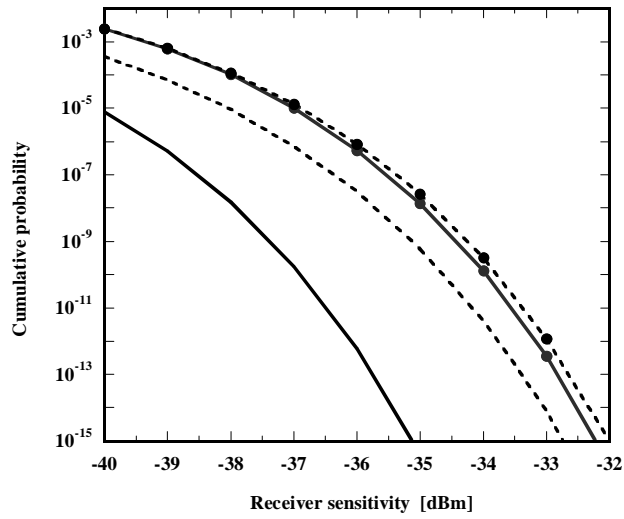

**Fig. 2**

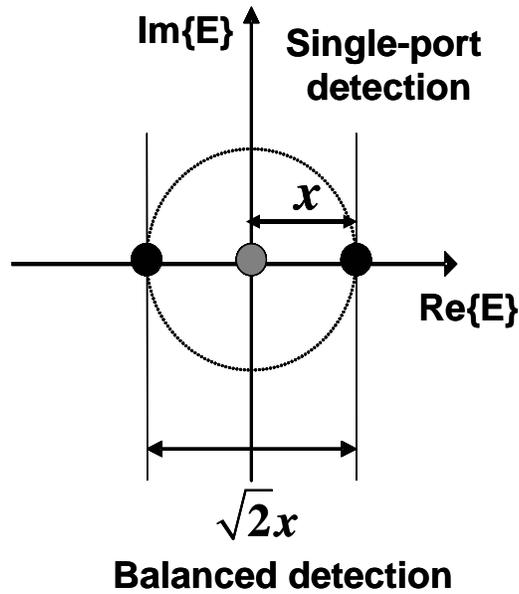

**Fig.3**